\documentclass[fleqn,twoside]{article}
\usepackage{espcrc2}

\title{Calculation of Nucleon Electromagnetic Form Factors}

\author{LHPC and SESAM Collaborations: D. B. Renner\address[MIT]{Center for
Theoretical Physics, Massachusetts
Institute of Technology, Cambridge,
MA 02139, USA}\thanks{Support provided through the U.S. Department of
Energy Cooperative Agreement DE-FC02-94ER40818},
R. Brower\address[BU]{Department of Physics, Boston
University, Boston, MA 02215, USA}, D. Dolgov\addressmark[MIT]$^\ast$,
N. Eicker\address[Wup]{Department of Physics, University of Wuppertal,
D-42097 Wuppertal, Germany}, Th. Lippert\addressmark[Wup],
J. W. Negele\addressmark[MIT]$^\ast$, A.
Pochinsky\addressmark[MIT]$^\ast$, K. Schilling\addressmark[Wup]
}

\bibliography{bib}

\begin{document}
\bibliographystyle{unsrt}
%
%
\begin{abstract}
The fomalism is developed to express nucleon matrix elements of the
electromagnetic
current  in terms of form factors consistent with the
translational, rotational, and
parity symmetries of a cubic lattice. We calculate the number of these
form factors
and show how appropriate linear combinations approach the  continuum
limit.

\end{abstract}
%
\maketitle
%
%

Electroweak form factors provide a precise experimental probe of the quark
substructure of hadrons, so it is of great interest to calculate them in
QCD from
first principles. Since  the only known way to solve
QCD is numerical solution on a discrete lattice, this work addresses the
decomposition
of matrix elements of the electromagnetic current into form factors
consistent with
cubic lattice symmetry and the relation of these form form factors to the
familiar ones associated with continuum Lorentz symmetry.

The breaking of continuum symmetries leads, in
general, to more form factors on the lattice than in the continuum. In a
practical
Monte Carlo calculation on the lattice, it is advantageous to calculate these
lattice form
factors since all  momenta related by a lattice symmetry can be combined to
obtain the highest statistical accuracy. Therefore, it is necessary to
determine the
appropriate combination of these form factors that approaches the physical
form factors in the continuum limit.

In this work, we consider the electromagnetic form factors of the nucleon.
The matrix element of the electromagnetic current in the continuum is
\begin{eqnarray*}
\lefteqn{ \langle p^\prime,s^\prime|J^\mu|p,s\rangle ~ = } \\
 & & \overline{u}(p^\prime,s^\prime)\left( \gamma^\mu F_1(q^2) + i \frac{
\sigma^{\mu \nu} q_\nu }{2m} F_2(q^2) \right) u(p,s)
\end{eqnarray*}
where $p$ and $s$ ($p^\prime$ and $s^\prime$) are the initial (final)
momentum
and spin z-component and $J_\mu$ is the electromagnetic current. This is
the most
general form consistent with continuum space-time symmetries and
conservation of
the electromagnetic current.  We seek an analogous result on the lattice.

It is easiest to consider
the following lattice matrix element between spin and helicity states:
\begin{eqnarray*}
\langle\vec{0},s| J_\mu |\vec{p},h\rangle.
\end{eqnarray*}
The final state has zero momentum and spin z-component $s$.  The initial
state has an arbitrary non-zero momentum $\vec{p}$ and helicity $h$.
The fact that helicity states do not mix under rotations simplifies our
analysis.
We now need to
construct the
most general form of this matrix element consistent with lattice spatial
symmetries.

To construct lattice form factors, we
decompose the non-zero momentum state into irreducible representations of the
lattice rotation group and apply the Wigner-Eckart theorem to write
the matrix element in terms of a set of reduced matrix elements.
These reduced matrix elements are the lattice form factors.
Since the final state is already irreducible and the current simply
consists of two irreducible
representations, $J_0$ and $J_{1,2,3}$, we only need the decomposition of the
non-zero momentum states and the Clebsch-Gordan coefficients.

To start, we consider the orbit of momentum $\vec{p}$ under the lattice rotation
group, that is,  all lattice rotations of $\vec{p}$.
This set of states forms a basis of a
generally reducible representation in the Hilbert space.  We block diagonalize
this representation,
producing a change of basis from linear to angular momentum states that may be
written as follows:
\begin{eqnarray*}
\lefteqn{ {|\vec{p}, h\rangle}_{C} = } \\
 & & \sum_{R=1}^{8} \sum_{\alpha=1}^{n_R} \sum_{s=1}^{d_R} A_{C,h}(\vec{p}
; R, \alpha, s ) {|R,\alpha,s\rangle}_{C,h}
\end{eqnarray*}
where $R$ labels the eight irreducible, double-valued representations
of the universal cover of the lattice rotation group (Ref. \cite{Mandula:ut},
\cite{Mandula:wb}),
$n_R$ is the number of occurrences of $R$, $\alpha$ labels each occurrence,
$d_R$ is the dimension of $R$, $s$ labels the basis states,
$C$ denotes a momentum class from which $p$ takes its values, and $h$ is
the helicity.  All the momentum
classes (except for irrelevant special cases at the edge of the Brillouin zone)
can be catalogued into the eight cases  shown in Table 1, of which only five
have essentially different $A$'s.  Note that all momenta are lattice momenta,
and physical momenta require a factor of $\pi/L$.
\begin{table}
\caption{ For each type of momentum class, where $N>M>L>0$, we
list the order of the class,
the number of $\frac{1}{2}$ and $\frac{3}{2}$ representations in
the decomposition, the number of $G$ and $F$ lattice form factors, and the
total number of lattice form factors. }
\begin{tabular}{ccccccc} \hline
class & order & $\frac{1}{2}$ & $\frac{3}{2}$ & $G$ & $F$ & $G+F$ \\ \hline
$[( 0, 0,  0 )]$ & 1  & 1 & 0 & 1 & 1 & 2 \\
$[( N, 0,  0 )]$ & 6  & 1 & 1 & 1 & 2 & 3 \\
$[( N, N,  N )]$ & 8  & 1 & 1 & 1 & 2 & 3 \\
$[( N, N,  0 )]$ & 12 & 1 & 2 & 1 & 3 & 4 \\
$[( N, M,  0 )]$ & 24 & 2 & 4 & 2 & 6 & 8 \\
$[( N, M,  M )]$ & 24 & 2 & 4 & 2 & 6 & 8 \\
$[( N, N,  M )]$ & 24 & 2 & 4 & 2 & 6 & 8 \\
$[( N, M,  L )]$ & 24 & 2 & 4 & 2 & 6 & 8 \\
$[( N, M, -L )]$ & 24 & 2 & 4 & 2 & 6 & 8 \\ \hline
\end{tabular}
\end{table}

The general results for matrix elements may be written as follows:
\pagebreak
\begin{eqnarray*}
\lefteqn{ \langle\vec{0},s|J_0{|\vec{p}, h\rangle}_{C} = } \\
&& \sum_{\alpha}^{n_\frac{1}{2}} A_{C,h}(\vec{p};\frac{1}{2},\alpha,s)
G^h_\alpha \\
\lefteqn{ \langle\vec{0},s|J_i{|\vec{p}, h\rangle}_{C} = } \\
&& \sum_{\alpha}^{n_\frac{1}{2}} B_{C,h}(\vec{p};1/2,\alpha,s,i)
F_{\frac{1}{2}\alpha}^h + \\
&& \sum_{\alpha}^{n_\frac{3}{2}} B_{C,h}(\vec{p};3/2,\alpha,s,i)
F_{\frac{3}{2}\alpha}^h \\
\lefteqn{ B_{C,h}(\vec{p};R,\alpha,s,i) = } \\
&& \sum_{s^\prime}^{d_R} A_{C,h}(\vec{p};R,\alpha,s^\prime)
C(1/2,s,1,i;R,s^\prime)
\end{eqnarray*}
where $C(1/2,s,1,i;R,s^\prime)$ denotes the Clebsch-Gordan coefficients
coupling
$\frac{1}{2}
\otimes 1$ to $R$. The states $\langle\vec{0},s|J_0$ transform as the $\frac{1}{2}$
representation and hence have overlap only with the $\frac{1}{2}$ component of
${|\vec{p}, h\rangle}_{C}$.  Therefore for each occurrence of the $\frac{1}{2}$
representation
in ${|\vec{p}, h\rangle}_{C}$ there is an undetermined reduced matrix element,
$G$. The states $\langle\vec{0},s|J_i$ transform
as the $\frac{1}{2} \otimes 1 = \frac{1}{2} \oplus \frac{3}{2}$
representation and hence overlap only with the
$\frac{1}{2}$ and $\frac{3}{2}$ pieces of ${|\vec{p}, h\rangle}_{C}$.  Similarly
for each occurrence of the $\frac{1}{2}$ or $\frac{3}{2}$ representation
in the decomposition there is an undetermined matrix element,
$F_\frac{1}{2}$ or $F_\frac{3}{2}$.

To illustrate the above results, we discuss two examples.
The first treats momenta for which the continuum form of the $J_0$
matrix
element is preserved on the lattice.  This occurs for several types of momentum
classes:
$[(0,0,0)]$, $[(N,0,0)]$, $[(N,N,N)]$, and $[(N,N,0)]$.  The result for
momenta of the type $[(N,0,0)]$ is
\begin{eqnarray*}
{ \langle\vec{0},s|J_0|\vec{p},h\rangle = }
\frac{1}{\sqrt{3}} \chi_s( h \hat{p} ) G^\mathrm{lat}_1
\end{eqnarray*}
where  $ \chi_s(  \hat{p} )$  denotes  a standard  spin one-half spinor
polarized  in
the $\hat{p}$ direction  and  $ G^\mathrm{lat}_1 $ has the following
continuum limit
\begin{eqnarray*}
\lefteqn{ G^\mathrm{lat}_1 \rightarrow \sqrt{3} \sqrt{ \frac{E+m}{2m}}
G^\mathrm{cont}_e .}
\end{eqnarray*}
Note that there are arbitrarily large lattice momenta for which the continuum
form is retained.

The second example treats momenta for which the $J_0$ matrix element has
an additional form factor.  This occurs for the remaining types of momentum
classes,
and we consider momenta of the type $[(N,M,0)]$.  Since there are two
independent
terms   $G^h_\alpha$,  we have taken a linear combination such that one term
is proportional to the spinor $\chi_s( h \hat{p} )$, with the result

\begin{eqnarray*}
\lefteqn{ \langle\vec{0},s|J_0|\vec{p},h\rangle = } \\
&& \frac{1}{\sqrt{12}} \chi_s( h \hat{p} ) G^\mathrm{lat}_1 +
\frac{1}{\sqrt{12}} \chi_s( -h \hat{p} ) G^\mathrm{lat}_2.
\end{eqnarray*}
Hence, the two form factors have the continuum limit
\begin{eqnarray*}
\lefteqn{ G^\mathrm{lat}_1 \rightarrow \sqrt{12} \sqrt{ \frac{E+m}{2m}}
G^\mathrm{cont}_e } \\
\lefteqn{ G^\mathrm{lat}_2 \rightarrow 0 }
\end{eqnarray*}
and we have succeeded in identifying the linear combination that corresponds to
continuum physics.

Note that on the cubic lattice, in addition to the $\frac{1}{2}$ representation
corresponding to the physical angular momentum  $\frac{1}{2}$ state in the
continuum, there exists a second  $\frac{1}{2}$ representation originating
from higher angular momentum continuum states that are imperfectly
represented on the cubic lattice. Since there is no physical matrix element in
the continuum connecting the  $\frac{1}{2}$ state to these higher angular
momentum states, it follows that the coefficient $G^\mathrm{lat}_2$ must
vanish in the continuum limit.

The relation between behavior on the lattice and in the continuum can also be
clarified
by considering the little group for each non-zero momentum.
The little group, or stabilizer,  is the sub-group of lattice rotations which leave the
momentum invariant.  Each irreducible representation of the little group leads
to an allowed helicity, as summarized in Table 2.  Just as the number of
distinct spins is restricted
on the lattice,
the number of allowed helicities is also limited and dependent upon momentum.
In the continuum $G^\mathrm{lat}_2$ would mediate
an $\frac{1}{2} \rightarrow -\frac{1}{2}$ transition, but on the lattice, for
this type of momentum class, $-\frac{1}{2}$ is isomorphic to $\frac{1}{2}$,
thus the lattice
does not have enough constraints to exclude $G^\mathrm{lat}_2$.
\begin{table}
\caption{ For each type of non-zero momentum class we list: the little
group, the allowed helicities,
and the helicity range. }
\begin{tabular}{cclc} \hline
class & LG & helicities & $\Delta h$ \\ \hline
$[( N, 0,  0 )]$ & $Z_8$ & $-\frac{3}{2}$, $-1$, $-\frac{1}{2}$, $0$,
$\frac{1}{2}$, $1$, $\frac{3}{2}$, $2$ & $4$ \\
$[( N, N,  N )]$ & $Z_6$ & $-1$, $-\frac{1}{2}$, $0$, $\frac{1}{2}$, $1$,
$\frac{3}{2}$ & $3$  \\
$[( N, N,  0 )]$ & $Z_4$ & $-\frac{1}{2}$, $0$, $\frac{1}{2}$, $1$ & $2$  \\
$[( N, M,  0 )]$ & $Z_2$ & $0$, $\frac{1}{2}$ & $1$  \\
$[( N, M,  M )]$ & $Z_2$ & $0$, $\frac{1}{2}$ & $1$  \\
$[( N, N,  M )]$ & $Z_2$ & $0$, $\frac{1}{2}$ & $1$  \\
$[( N, M,  L )]$ & $Z_2$ & $0$, $\frac{1}{2}$ & $1$  \\
$[( N, M, -L )]$ & $Z_2$ & $0$, $\frac{1}{2}$ & $1$  \\ \hline
\end{tabular}
\end{table}

The previous example demonstrates a general feature: the order of the
little group determines the extent to which
the continuum form is identically adhered to by the lattice.  A larger
little group ensures
sufficient symmetry to enforce the continuum results, whereas a smaller
little group
lacks the constraints to prevent additional form factors.

To summarize, the lattice breaks continuum symmetries, and hence the number
of lattice form factors increases.
The remaining symmetries vary with momentum, consequently, there are
momenta which admit one to six
extra form factors.
We can understand the presence of these additional form factors in terms
of the
degraded range of helicities for each momentum.
The explicit decomposition of lattice matrix elements allows us to distinguish
these unphysical lattice form factors.
Our future work will apply this analysis to the SESAM
configurations
and compare it with the conventional analysis using continuum expressions.

%
%
%

%
%
%
%

\begin{thebibliography}{boo}
%
\bibitem{Mandula:ut}
J.~E.~Mandula, G.~Zweig and J.~Govaerts,
Nucl.\ Phys.\ B {\bf 228}, 91 (1983).
%
\bibitem{Mandula:wb}
J.~E.~Mandula and E.~Shpiz,
Nucl.\ Phys.\ B {\bf 232}, 180 (1984).
%
\end{thebibliography}
\end{document}